\documentclass{ifacconf}

\usepackage{amstext}
\usepackage{amsfonts}
\usepackage{amssymb}
\usepackage{epsfig}
\usepackage{graphicx}
\begin{document}
\begin{frontmatter}

\title{Time-delayed feedback control of delay-coupled neurosystems and lasers
} 
\author[]{Philipp H\"ovel, Markus Dahlem, Thomas Dahms, Gerald Hiller,} 
\author[Fifth]{and Eckehard Sch\"oll}

\address[Fifth]{Institut f{\"u}r Theoretische Physik, Technische
Universit{\"a}t Berlin, 10623 Berlin, Germany  (e-mail: schoell@physik.tu-berlin.de)}

\begin{abstract} 
We discuss applications of time-delayed feedback control to delay-coupled neural systems and lasers, 
in the framework of the FitzHugh-Nagumo neuron model and the Lang-Kobayashi laser model, respectively. In the context of
neural systems, we will 
point out some complex scenarios of synchronized in-phase or antiphase oscillations, bursting patterns, or amplitude
death, induced by delayed coupling in combination with delayed self-feedback in simple network motifs. For optical
systems, we will show that multiple time-delayed feedback, realized by a Fabry-Perot resonator coupled to the laser, 
provides a valuable tool for the suppression of unwanted intensity pulsations, and leads to stable continuous-wave
operation.
\end{abstract}


\end{frontmatter}
Over the past decade control of unstable and chaotic states has evolved into a central issue in applied nonlinear science
\citep{SCH07}. Various methods of control have been developed since the ground-breaking work of
\citet{OTT90}. One scheme where the control force is constructed from time-delayed signals \citep{PYR92} has turned out
to be very robust and universal to apply, and easy to implement experimentally. In this time-delayed feedback control
the control signal is built from the difference $s(t)-s(t-\tau )$ between the present and an earlier value of an
appropriate system variable $s$. It is noninvasive since the control forces vanish if the target state (a periodic
state of period $\tau$ or a steady state) is reached. Thus the unstable states themselves of the uncontrolled system are
not changed, but only  their neighbourhood is adjusted such that neighbouring trajectories converge to it,
i.e., the control forces act only if the system deviates from the state to be stabilized. 

This paper is organized as follows: In Sec.~\ref{sec:neuro}, we discuss the dynamics of delay-coupled
neurons and investigate the period of oscillations which can be induced for sufficiently large delay and
coupling strength. The additional application of time-delayed self-feedback leads to complex scenarios of synchronized
in-phase or 
antiphase oscillations, bursting patterns, or amplitude death. The stabilization of steady states in optical systems is
studied in Sec.~\ref{sec:laser} in the framework of a modified Lang-Kobayashi model of a semiconductor laser where the
time-delayed feedback is realized by a Fabry-Perot (FP) resonator.

\section{Neurosystems}
\label{sec:neuro}
In order to grasp the complicated interaction between billions of neurons in large neural networks, those are often
lumped into groups of neural populations each of which can be represented as an effective excitable element that is
mutually coupled to the other elements \citep{ROS04,POP05}. In this sense the simplest model which may reveal features 
of interacting neurons consists of two coupled neural oscillators. Each of these will be represented by a simplified
FitzHugh-Nagumo system \citep{FIT61,NAG62}, which is often used as a paradigmatic generic model for neurons, or more
generally, excitable systems \citep{LIN04}. Here we use two identical FitzHugh-Nagumo systems with parameters 
corresponding to the excitable regime.
 
We consider the simultaneous action of delayed coupling and delayed self-feedback.  Here we choose to
apply the self-feedback term symmetrically to both activator equations, but other feedback schemes
are also possible. The equations of the system are given by
\begin{eqnarray}
  \epsilon \dot{u}_1 &=& u_1 - \frac{u_1^3}{3} - v_1+ C[u_2(t-\tau)-u_1(t)]\nonumber\\
  \label{eq:FHN_u1}
  &\,&+K[u_1(t-\tau_K)-u_1(t)]\\
  \label{eq:FHN_v1}
  \dot{v}_1 &=& u_1 + a \\
  \epsilon \dot{u}_2 &=& u_2 - \frac{u_2^3}{3} - v_2 + C[u_1(t-\tau)-u_2(t)]\nonumber\\
  \label{eq:FHN_u2}
  &\,&+K[u_2(t-\tau_K)-u_2(t)]\\
  \label{eq:FHN_v2}
  \dot{v}_2 &=& u_2 + a,
\end{eqnarray}
where subsystems Eqs.~(\ref{eq:FHN_u1})-(\ref{eq:FHN_v1}) and (\ref{eq:FHN_u2})-(\ref{eq:FHN_v2}) represent two
different neurons (or neuron populations), $u_i$ ($i=1,2$) describing the activator (e.g., transmembrane voltages) 
and $v_i$ modelling the inhibitor (e.g., electrical conductances of the relevant ion currents across the
respective membranes).  Here $a$ is a bifurcation parameter whose value defines whether the system is excitable ($a>1$)
or demonstrates periodic firing, i.e., autonomous oscillations ($a<1$), $\epsilon$ is positive parameter that is usually
chosen to be much smaller than unity, corresponding to fast activator variables $u_1$, $u_2$, and slow inhibitor
variables $v_1$, $v_2$. 

\begin{figure}[th]
\begin{center}
	\includegraphics[width=0.75\linewidth]{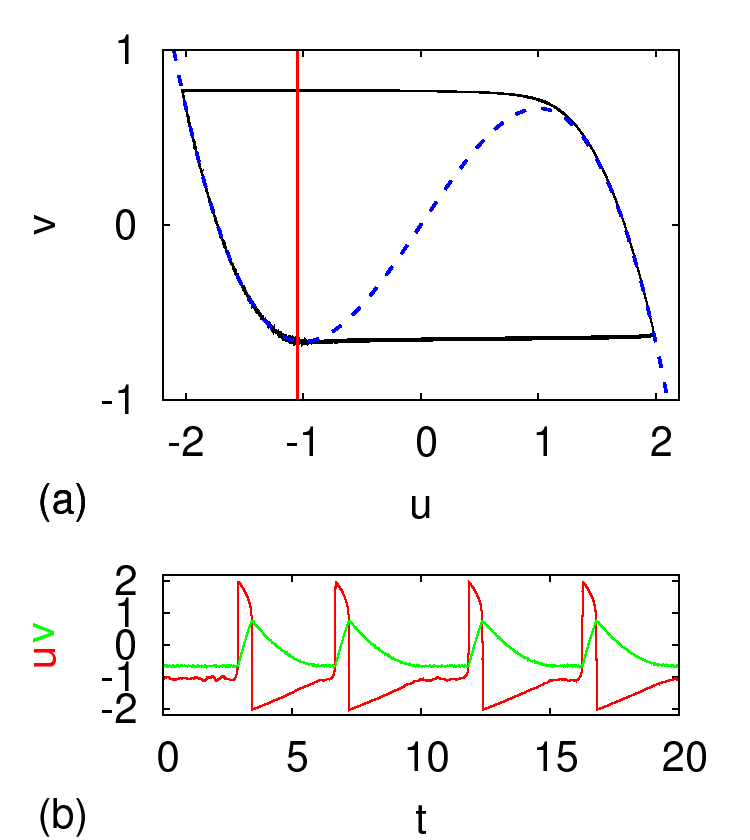}
\end{center}
\caption{Panel (a): Phase portrait with nullcline of the single FitzHugh-Nagumo system without time-delayed feedback.
Panel (b) Time series of the activator $u$ and inhibitor $v$ as red and green curves, respectively. Parameters:
$\epsilon=0.005$, $a=1.05$, and noise $D=0.02$.}
\label{fig:neuro_1FHN_nullcline}
\end{figure}

Let us illustrate the dynamics of a single neuron model by considering an uncoupled subsystem Eqs.~(\ref{eq:FHN_u1}) and
(\ref{eq:FHN_v1}) ($C=0$, $K=0$) under the influence of noise. This is realized by small random fluctations modeled
as Gaussian white noise $D\zeta(t)$ with noise intensity $D$ applied to Eq.~(\ref{eq:FHN_v1}). We fix $D=0.02$, and also
set $a=1.05$, $\epsilon=0.005$. In Fig.~\ref{fig:neuro_1FHN_nullcline}(a) the dashed blue and solid red curves show the
nullclines of Eq.~(\ref{eq:FHN_u1}) and (\ref{eq:FHN_v1}), respectively, which intersect at a fixed point. The phase
point that is initially placed at the fixed point stays in its close vicinity if the applied random perturbation remains
small. However, if the perturbation is larger than some threshold value, the phase point makes a large excursion in the
phase space before returning to the vicinity of the fixed point again. In Fig.~\ref{fig:neuro_1FHN_nullcline}(a) the
black solid line illustrates such a phase trajectory and in Fig.~\ref{fig:neuro_1FHN_nullcline}(b) realizations of $u_1$ and
$v_1$ time series from Eqs.~(\ref{eq:FHN_u1}) and (\ref{eq:FHN_v1}) are shown. The motion of the phase point consists of
two stages: an activation time during which the system waits for a sufficiently large perturbation before it can make an
excursion, and the excursion itself. The excursion time is almost completely defined by the deterministic properties of
the system and is hardly influenced by noise.

The synaptic coupling between two neurons in Eqs.~(\ref{eq:FHN_u1})-(\ref{eq:FHN_v2}) is modelled as a diffusive
coupling considered for simplicity to be symmetric \citep{LIL94,PIN00,DEM01}. More general delayed couplings are
considered in \citep{BUR03}. The coupling strength $C$ summarizes how information is distributed between neurons.  The
mutual delay $\tau$ in the coupling is motivated by the propagation delay of action potentials between the two neurons
$u_1$ and $u_2$.

Besides the delayed coupling we also consider delayed self-feedback in the form suggested by Pyragas \citep{PYR92},
where the difference $s(t)-s(t-\tau_K)$ of a system variable $s$, e.g., activator or inhibitor, at time $t$ and at a
delayed time $t-\tau_K$, multiplied by some control amplitude $K$, is coupled back into the same system. Such feedback
loops might arise naturally in neural systems, e.g., due to neurovascular couplings that has a characteristic latency,
or due to finite propagation speed along cyclic connections within a neuron sub-population, or they could be realized by
external feedback loops as part of a therapeutical measure, as proposed by \citet{POP05}. 
This feedback scheme is simple to implement and quite robust.
One distinct advantage of this method is its noninvasiveness, i.e., in the ideal deterministic limit the
control force vanishes on the target orbit, which may be a steady state or a periodic oscillation of period $\tau$. In
case of noisy dynamics the control force, of course, does not vanish but still remains small, compared to other common
control techniques using external periodic signals, for instance, in deep-brain stimulation to suppress neural synchrony
in Parkinson's disease \citep{TAS02}.

By a linear stability analysis, it can be shown that the fixed point remains stable for all values of $K$ and $\tau_K$
in case of $a>1$, as without self-feedback \citep{SCH08}. Redefining $\xi=1-a^2-C-K$, one obtains the factorized
characteristic equation
\begin{equation}
1- \xi \lambda+\epsilon \lambda^2 = \lambda K e^{-\lambda \tau_K} \pm \lambda C e^{-\lambda \tau }
\label{char_eq_K}
\end{equation}
Substituting the Hopf condition $\lambda=i \omega$ and separating into real and imaginary parts yields for the 
imaginary part
\begin{equation}
-\xi =K \cos(\omega \tau_K) \pm C \cos(\omega \tau)
\end{equation}
This equation has no solution for $a>1$ since $|\xi|=a^2-1+C+K>C+K$.

\begin{figure}[ht]
	\begin{center}
		\includegraphics[width=0.9\linewidth]{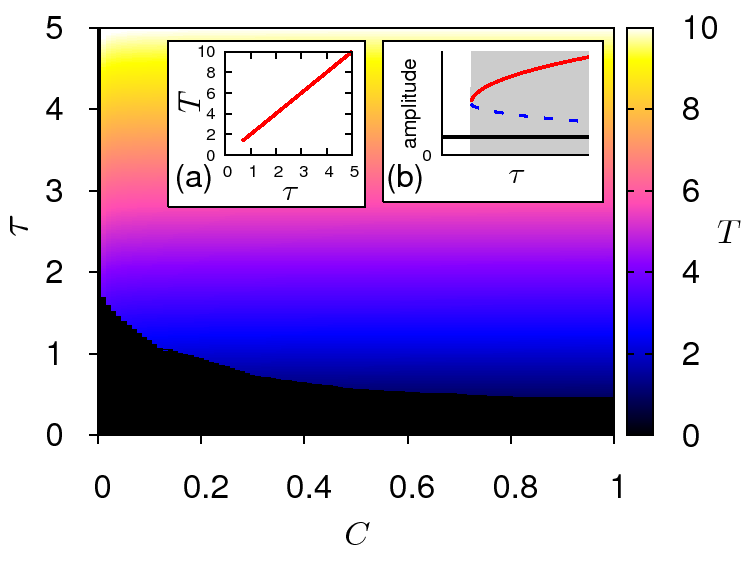}
	\end{center}
\caption{Regime of oscillations in the $(C,\tau)$ parameter plane for initial conditions corresponding to
single-pulse-excitation in one system. The oscillation period $T$ is color coded. The transition between black and color
marks the bifurcation line.  Inset (a) shows the oscillation period vs. $\tau$ in a cut at $C=0.8$. Inset (b): schematic
plot of the saddle-node bifurcation of a stable (red solid line) and unstable (blue dashed) limit cycle. The maximum
oscillation amplitude is plotted vs. the delay time $\tau$ and the stable fixed point is plotted as a solid black line.
The grey background marks the bistable region. Parameters: $a=1.05$, $\epsilon=0.01$, $K=0$  \citep{SCH08}.}
\label{fig:hauplot} 
\end{figure} 

In Fig.~\ref{fig:hauplot} the regime of oscillations is shown in the parameter plane of the coupling strength $C$ and
coupling delay $\tau$. The oscillation period is color coded. The boundary of this colored region is given by
the minimum coupling delay $\tau_{min}$ as a function of $C$. For large coupling strength, $\tau_{min}$ is almost
independent of $C$; with decreasing $C$ it sharply increases, and at some small minimum $C$ no oscillations exist at
all. At the boundary, the oscillation sets in with finite frequency and amplitude as can be seen in the insets of
Fig.~\ref{fig:hauplot} which show a cut of the parameter plane at $C=0.8$. The oscillation period increases linearly
with $\tau$.  The mechanism that  generates the oscillation is a saddle-node bifurcation \index{saddle-node bifurcation}
of limit cycles (see inset (b) of Fig.~\ref{fig:hauplot}), creating a pair of a stable and an unstable limit cycle. 
The unstable limit cycle separates the two attractor basins of the stable limit cycle and the stable fixed point.

\begin{figure}[ht]
	\begin{center}
		\includegraphics[width=0.975\linewidth]{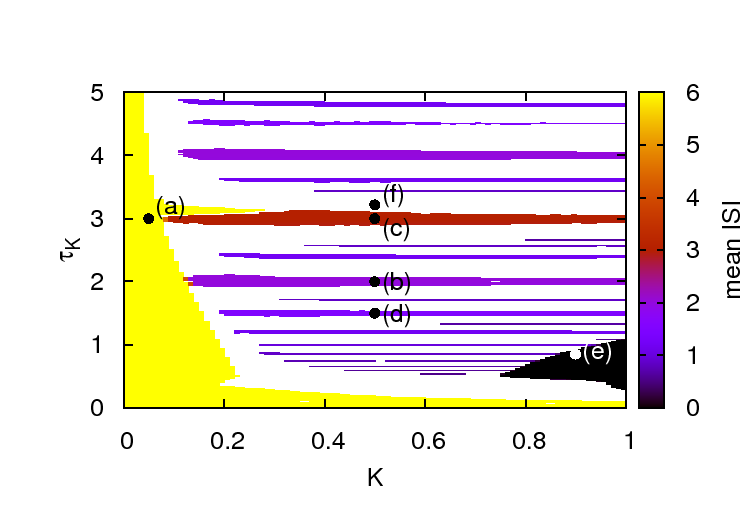}
	\end{center}
\caption{Influence of delayed self-feedback upon coupled oscillations. The mean interspike interval (ISI) is color coded
in the control parameter plane of the self-feedback gain $K$ and delay $\tau_K$. White areas mark regimes of irregular 
oscillations where the ISI variance becomes large ($>0.01$). Time series corresponding to points (a)-(f) are shown in 
Fig.~\ref{fig:controlcases}. Other parameters:  $a=1.3$, $\epsilon=0.01$, $C=0.5$, $\tau=3$ \citep{SCH08}.}
\label{fig:isiplane} 
\end{figure}

\begin{figure}[ht]
	\begin{center}
		\includegraphics[width=0.975\linewidth]{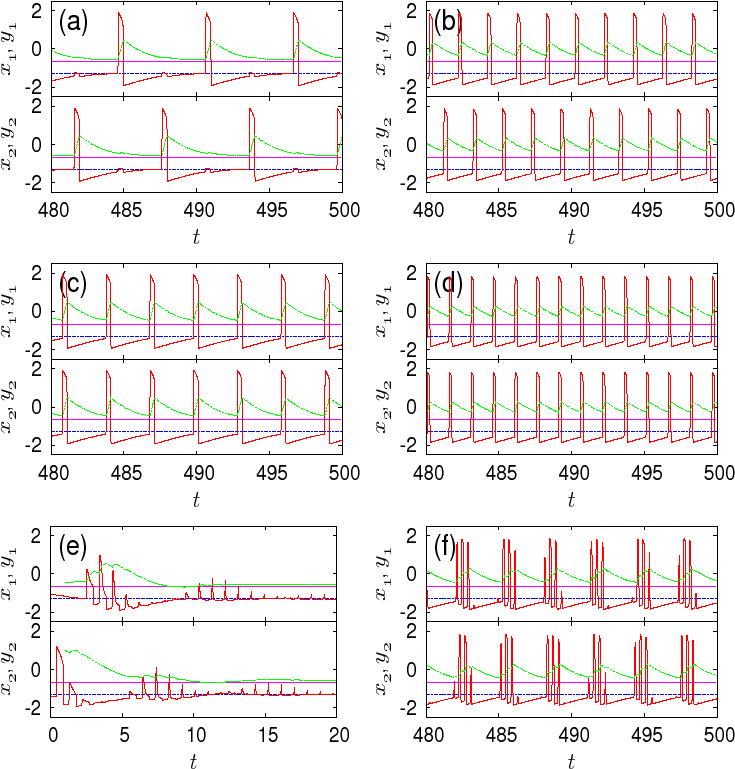}
	\end{center}
\caption{Different modes of oscillation corresponding to different self-feedback parameters $K$, $\tau$
(red solid lines: activators $u_i(t)$, green solid lines: inhibitors $v_i(t)$). (a), (b): Antiphase oscillations for (a)
$K=0.05, \tau_K=3$ (period $T=6$) and  (b) $K=0.5, \tau_K=2$ ($T=2$); (c), (d): In-phase oscillations for (c) $K=0.5,
\tau_K=3$ (period $T=3$) and (d) $K=0.5, \tau_K=1.5$ ($T=1.5$); (e): Oscillator death for $K=0.9, \tau_K=0.9$;
(f): Bursting pattern for $K=0.5, \tau_K=3.2$. Other parameters:  $a=1.3$, $\epsilon=0.01$, $C=0.5, \tau=3$
\citep{SCH08}.}
\label{fig:controlcases}
\end{figure}

The adopted form of control allows for the synchronization of the two cells not only for identical values of 
$\tau$ and $\tau_K$, but generates an intricate pattern of synchronization islands or stripes in the ($\tau, \tau_K$) 
control parameter
plane (Fig.~\ref{fig:isiplane}) corresponding to single-spike in-phase and antiphase oscillations with constant
interspike intervals, see also Fig.~\ref{fig:controlcases}(a)-(d). Further, for adequately chosen parameter sets of
coupling and self-feedback control, we observe effects such as bursting patterns Fig.~\ref{fig:controlcases}(f) and
oscillator death Fig.~\ref{fig:controlcases}(e). In addition to these effects, there exists a control parameter regime
in which the self-feedback has no effect on the oscillation periods (shaded yellow). 

Fig. \ref{fig:isiplane} shows the control parameter plane for coupling parameters of the uncontrolled system 
in the oscillatory regime ($C=0.5$ and $\tau=3$).  We observere three principal regimes: (i) Control has no effect on
the oscillation period (yellow), although the form of the stable limit cycle is slightly altered (Fig.
\ref{fig:controlcases}(a)). (ii) Islands of in-phase and antiphase synchronization (color coded, see
Fig.~\ref{fig:controlcases} (b)-(d)).(iii) Oscillator death (black) Fig.~ \ref{fig:controlcases} (e)). 

\section{Lasers}
\label{sec:laser}
Semiconductor lasers with external optical feedback from a mirror can be described by the Lang-Kobayashi (LK) model
\citep{LAN80b}. In dimensionless form, it consists of two differential equations for the slowly varying amplitude
(envelope) $E(t)$ of the complex electric field and the reduced carrier density (inversion) $n(t)$.

Here we consider a modification of the LK equations appropriate for multisection semiconductor lasers with
an internal passive dispersive reflector \citep{TRO00}. This is modeled by a gain function $k(n)$ depending upon the
internal dispersive feedback from the Bragg grating. Such a laser structure allows for more complex dynamic behavior
including self-sustained relaxation oscillations (intensity pulsations) generated by Hopf bifurcations, as has been
shown in the framework of traveling wave laser models \citep{BAU04,SCH06a}. We are interested in the regime 
above a supercritical Hopf bifurcation where the fixed point in the uncontrolled system is unstable. Combining the rate
equation for the carrier density from Ref.~\citep{TRO00} with the rate equation for the complex electric field, we
obtain the following form of modified LK equations:
\begin{eqnarray}
  \label{eqn:lk_tronciu}
  \frac{d E}{d t} & = & \frac{T}{2} \left( 1+ \imath \alpha \right) n E - E_{b}(t) ,\\
  \frac{d n}{d t} & = & I -n- (1+n) k(n) \left| E \right|^{2},
\end{eqnarray}
where $\alpha$ denotes the linewidth enhancement factor, $I$ is the reduced excess injection current, $T$ is the time
scale ratio of the carrier lifetime $\tau_c$ and the photon lifetime $\tau_p$, and $E_{b}(t)$ denotes the feedback term,
which will be described in detail later.

The function $k(n)$, which models the internal dispersive feedback, is chosen as a Lorentzian, as proposed by
\citet{TRO00}:
\begin{equation}
k(n)=k_{0} + \frac{A W^{2}}{4\left(n-n_{0}\right)^{2}+W^{2}} ,
\label{eqn:lk_kvonn}
\end{equation}
where $A$ denotes the height, $W$ is the width, and $n_{0}$ is the position of the resonance. 
The parameter $k_{0}$ is chosen such that $k(0)=1$ at the laser threshold. 
Througout the following we will use the parameters $A=1$, $W=0.02$, and $n_0 = -0.034$.

\begin{figure}[ht]
	\begin{center}
		\includegraphics[width=\linewidth]{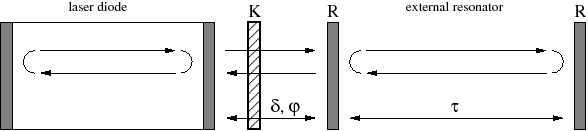}
	\end{center}
\caption{Schematic diagram of a semiconductor laser with resonant feedback from a Fabry-Perot resonator. 
$K$ denotes an attenuator, $R$ is related to the mirror reflectivity of the external resonator, $\tau$ is 
the round trip time of the resonator, $\varphi$ and $\delta$ are phase shift and latency time due to the distance
between laser and resonator \citep{DAH08b}.}
  \label{fig:laser}
\end{figure}

In Eqs.~(\ref{eqn:lk_tronciu}), the feedback term $E_{b}(t)$ has not been specified yet. In the following, 
we introduce the feedback term such that it models a Fabry-Perot resonator. As opposed to the original 
LK model where only a single external mirror is considered, we take an external FP resonator with multiple reflections
into account: 
\begin{eqnarray}
 E_{b}(t) & = & \kappa \sum_{m=0}^{\infty} R^{m} \left[ E(t-\delta-m \tau) - E(t-\delta-(m+1)\tau)
\right] \nonumber \\
 \label{eqn:lk_fp_feedback}
 & = & \kappa
 \left[ E(t-\delta) - E(t-\delta-\tau) \right] + R E_{b}(t-\tau).
\end{eqnarray}
$\tau$ is the delay time (cavity round trip time), $R$ is a memory parameter (mirror reflectivity), $\delta$ denotes the
latency time originating from a single round trip between the laser and the resonator (see Fig.~\ref{fig:laser}), and
$\kappa=K e^{-i\varphi}$ is the complex feedback gain with amplitude $K$ and the feedback phase $\varphi$ which results
from the associated optical phase shift. Throughout this work we use resonant feedback from the FP resonator.

The latency time $\delta$, i.e., the propagation time between the laser and the FP, is correlated to the phase
$\varphi$ by the relation $\varphi=\Omega_{0}\delta$, where $\Omega_{0}$ is the frequency of the emitted light.
However, we consider the two parameters $\varphi$ and $\delta$ as independent variables because the phase $\varphi$ can
be tuned by subwavelength changes of the separation between laser and FP resonator, on which scale the slowly varying
amplitude $E$, which depends upon $\delta$, does not change. The effect of latency in time-delayed feedback was already
studied in a general context in Refs.~\citep{JUS99b,HOE03,HOE05,DAH07}.

To investigate the stability of the cw laser emission, we will perform a linear stability analysis of the lasing state
is located at $(n=0,E=\sqrt{I} e^{\imath \psi})$ where the phase of the electric field can be arbitrarily fixed to
$\psi=0$ (solitary laser mode).

Using the abbreviations $E(t) = \sqrt{T} \left[ \Omega_{0} + x(t) + \imath y(t) \right]$, $\Omega_{0} =
\sqrt{I/T}$ and $\Gamma = T^{-1} \left[ 1 + I \left( 1 + \left.\frac{d k}{d n}\right|_{n=0}\right) \right]$
with real-valued $x$ and $y$, the fixed point is located at $(n=0,x=0,y=0)$. An exponential ansatz $\exp{(\Lambda t)}$
for all three variables $x$, $y$, and $n$
leads to the characteristic equation 
\begin{eqnarray}
 \label{eqn:EIGENVALUES_LK}
 0 & = & \left( 2 \Gamma + \Lambda \right) \left[ \left( K e^{-\Lambda\delta}\frac{1-e^{-\Lambda \tau}}{1- R e^{-\Lambda
\tau}} \right)^{2} + \Lambda^{2} \right.\\
 &\,& \left.+ 2 \Lambda K e^{-\Lambda\delta}\frac{1-e^{-\Lambda \tau}}{1- R e^{-\Lambda
\tau}} \cos{\varphi }\right] + 4 \Omega_{0}^{2} \left( \Lambda \right.\nonumber\\
 &\,&  \left.+ K e^{-\Lambda\delta}\frac{1-e^{-\Lambda \tau}}{1- R e^{-\Lambda \tau}}
\cos{\varphi} + \alpha K e^{-\Lambda\delta}\frac{1-e^{-\Lambda \tau}}{1- R e^{-\Lambda \tau}} \sin{\varphi}
\right).\nonumber
\end{eqnarray}
In our simulations, we use the following parameters, which were chosen close to the values by \citet{TRO00}:
$\Omega_{0} =0.06$, $\alpha =5$, $\Gamma =-0.01$ (corresponding to $T=500$,  $I = 1.8$,  $A = 1$, $W=0.02$, $n_{0} =
-0.034$, and $k_{0} = 0.993$). Thus the intrinsic period of the uncontrolled unstable focus is $T_{0} \approx
\pi/\Omega_{0} \approx 52$.

The stability of the lasing fixed point is given by Eq.~(\ref{eqn:EIGENVALUES_LK}). We solved this equation via Newton's
method. Since the transcendental equation has an infinite number of roots, we scanned the complex plane as 
initial conditions of the root-finding algorithm to locate the eigenvalue with the largest real part. The parameter
space consisting of $K$, $\tau$, $R$, $\varphi$, and $\delta$ is five-dimensional for fixed $\Gamma$ and $\Omega_{0}$.
To visualize the domain of control, we consider two-dimensional sections of this parameter space.

\begin{figure}[ht]
	\begin{center}
		\includegraphics[width=0.975\linewidth]{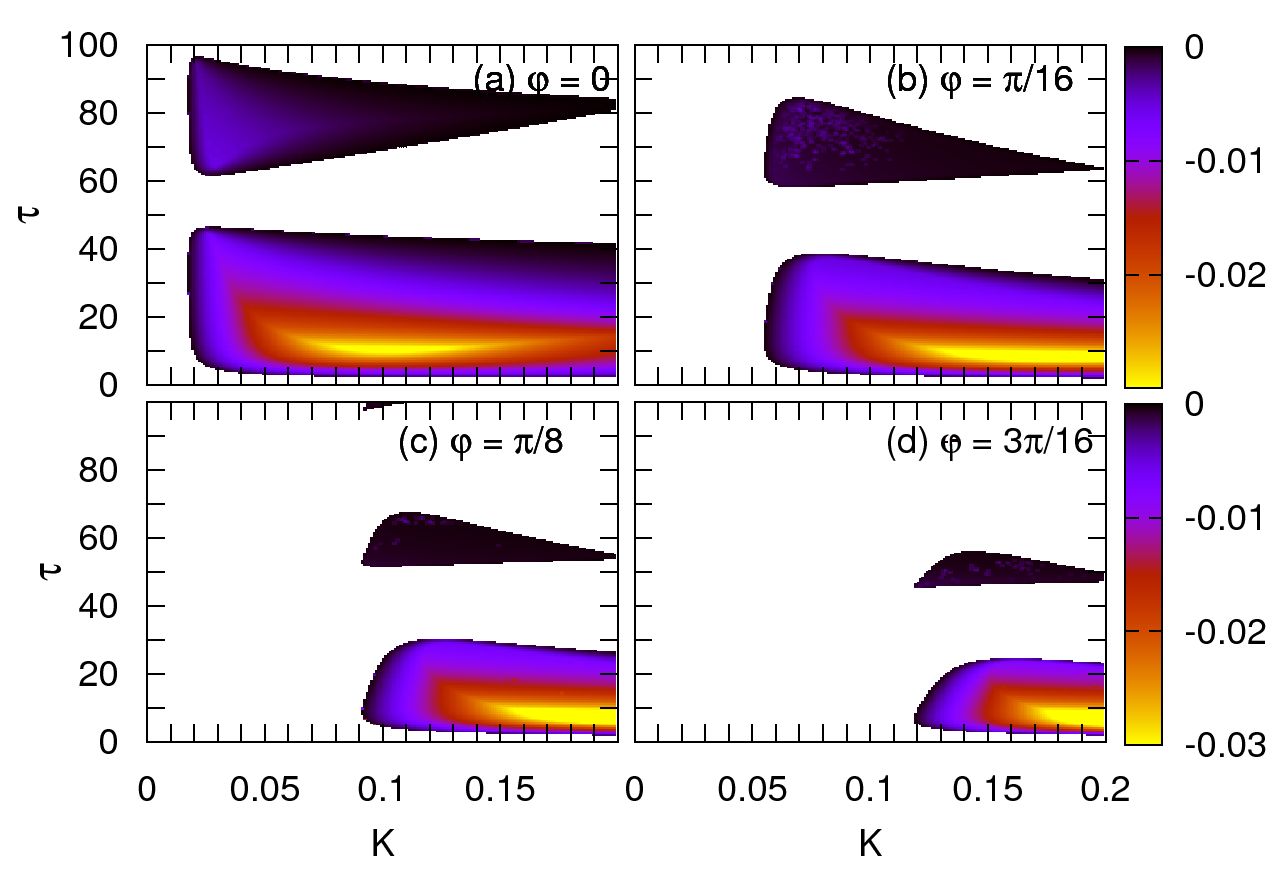}
	\end{center}
\caption{Domain of control according to Eq.~(\ref{eqn:EIGENVALUES_LK}) in the $(K,\tau)$-plane
for different values of $\varphi$. The color code denotes the largest real part $\mathrm{Re}(\Lambda)$ of
the eigenvalues $\Lambda$; only negative values are plotted. Panels (a), (b), (c), and (d) correspond to $\varphi=0$,
$\pi/16$, $\pi/8$, and $3\pi/16$, respectively. Other parameters: $\Gamma=-0.01$, $\Omega_{0}=0.06$, $\alpha=5$,
$R=0.7$, and $\delta=0$  \citep{DAH08b}.}
\label{fig:lk_k_tau_domain_phi}
\end{figure}
In Fig.~\ref{fig:lk_k_tau_domain_phi}, the domain of control is shown in the $(K,\tau)$-plane for different values of
the phase: $\varphi=0$, $\pi/16$, $\pi/8$, and $3\pi/16$ in panels (a), (b), (c), and (d), respectively. The
color code denotes the largest real part of the eigenvalues and is therefore a measure of stability. Note that
only values of $\mathrm{Re}(\Lambda)<0$ are plotted, thus the shaded regions correspond to a stable lasing fixed point,
i.e., a stable cw output. The control domains form tongues separated by (white) regions of no control around $\tau=n
T_{0}$ with $n$ integer, just as in the generic model studied in Refs.~\citep{HOE05,DAH07}. It can be seen that the
domain of control shrinks with increasing phase.
Here, the domains of control are cut off by boundaries from the upper left and right for increasing phase, leading to
overall smaller regions of stability. The tongues of stabilization are also slightly distorted towards smaller values of
$\tau$. Additionally, in this picture, the regions of optimum stability, denoted by yellow color, are shifted towards
larger values of the feedback gain $K$.

\begin{figure}[ht]
	\begin{center}
		\includegraphics[width=0.975\linewidth]{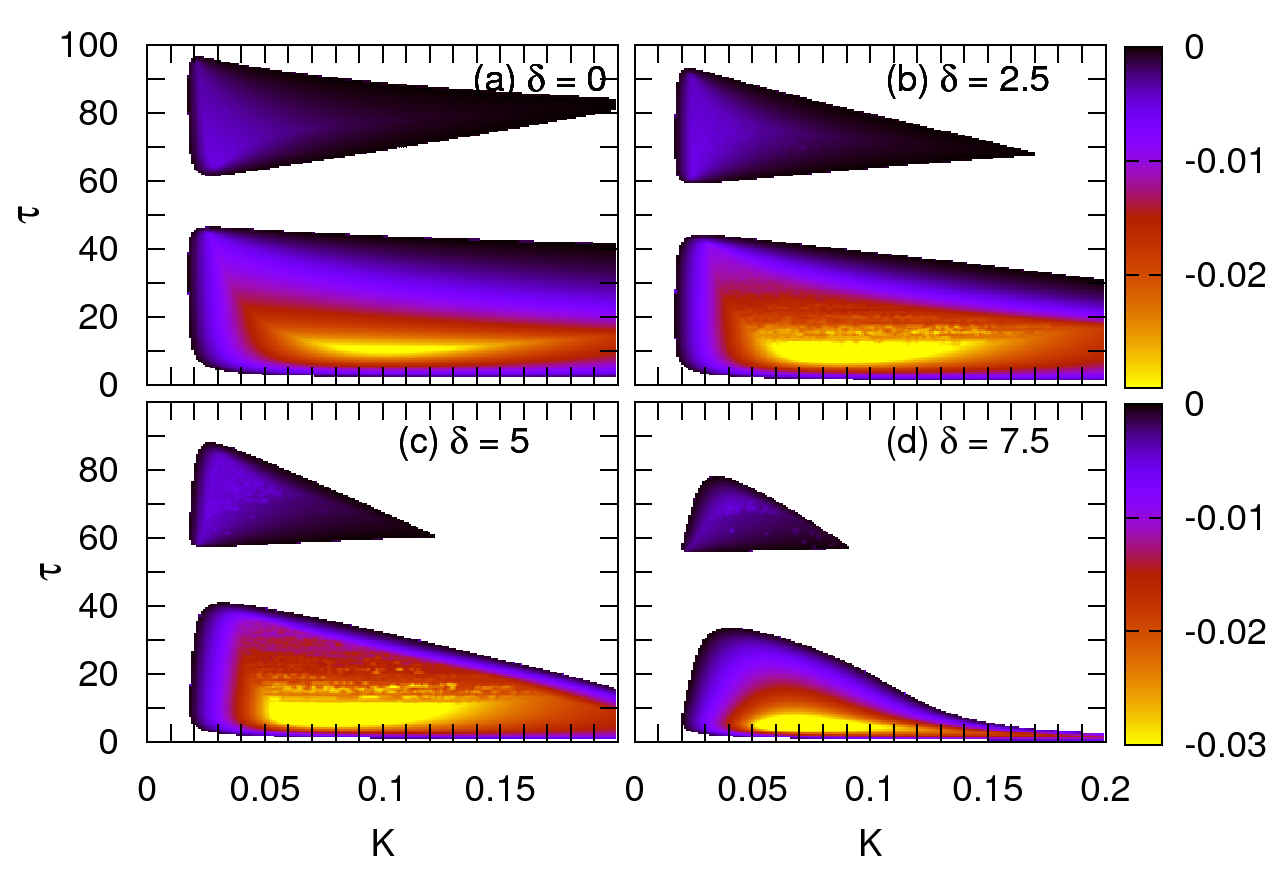}
	\end{center}
\caption{Domain of control in the $(K,\tau)$-plane for different values of $\delta$ and fixed $\varphi=0$. The color
code denotes the largest real part $\mathrm{Re}(\Lambda)$ of the eigenvalues $\Lambda$; only negative values are
plotted. Panels (a), (b), (c), and (d) correspond to $\delta=0$, $2.5$, $5$, and $7.5$, respectively. Other parameters
as in Fig.~\ref{fig:lk_k_tau_domain_phi} \citep{DAH08b}.}
\label{fig:lk_k_tau_domain_latency}
\end{figure}
Next, we will investigate the role of the latency time in the $(K,\tau)$-plane. In
Fig.~\ref{fig:lk_k_tau_domain_latency},
the domain of control in the $(K,\tau)$-plane is depicted for different values of the latency time, i.e., $\delta=0$,
$2.5$, $5$, and $7.5$, and fixed $\varphi=0$. For larger latency times, the domains of control shrink and it can also be
observed that they are bent down towards smaller values of the time delay $\tau$. 
Note that the regions of optimum stability, denoted by yellow color, are only slightly affected by the change
of the latency time.

\begin{figure}[ht]
	\begin{center}
		\includegraphics[width=0.975\linewidth]{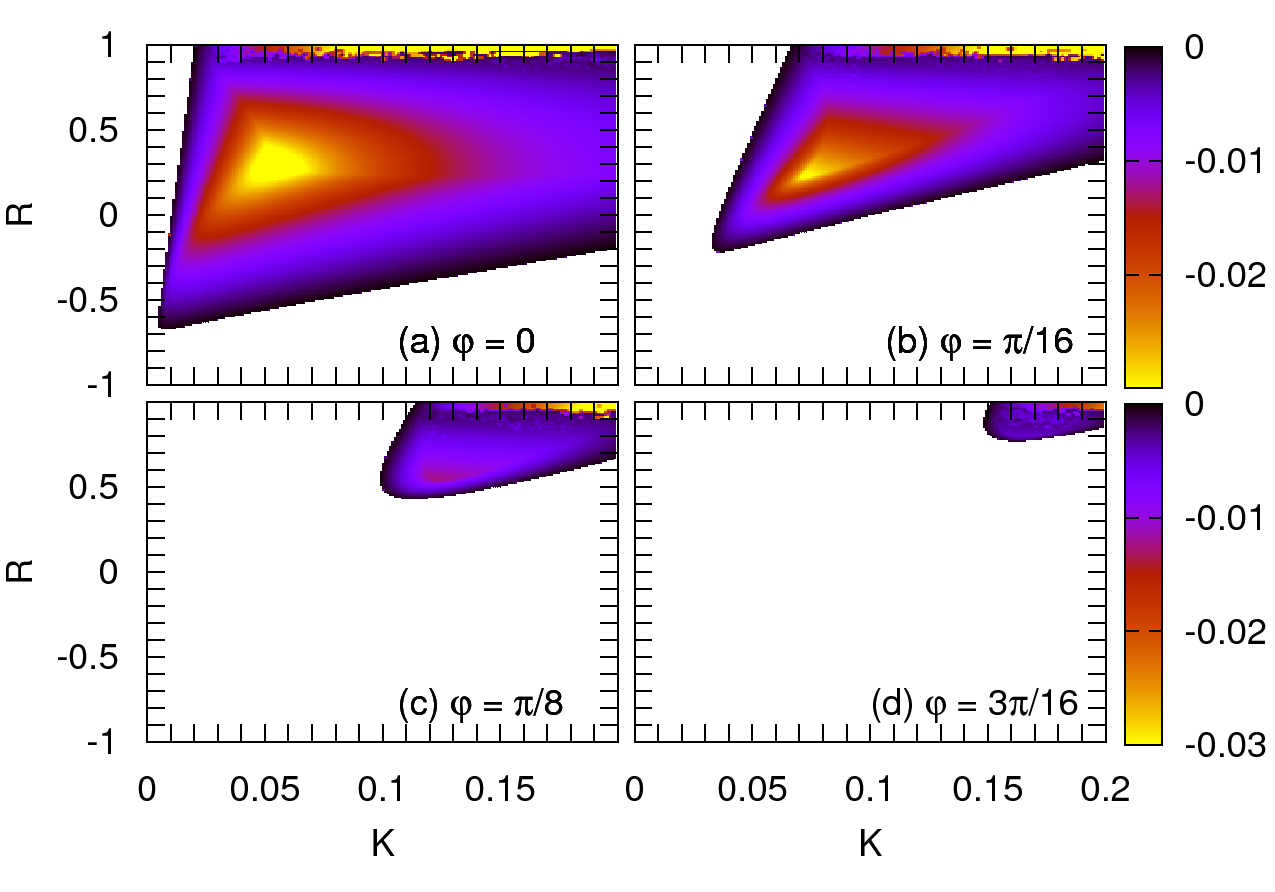}
	\end{center}
\caption{Domain of control in the $(K,R)$-plane for different values of $\varphi$ and fixed optimum time delay
$\tau=26$. The color code denotes the largest real part $\mathrm{Re}(\Lambda)$ of the eigenvalues $\Lambda$; only
negative values are plotted. Panels (a), (b), (c), and (d) correspond to $\varphi=0$, $\pi/16$, $\pi/8$, and $3\pi/16$,
respectively.  Other parameters as in Fig.~\ref{fig:lk_k_tau_domain_phi}  \citep{DAH08b}.}
\label{fig:lk_k_r_domain_phi}
\end{figure}

All figures shown here were obtained for a fixed value of the memory parameter $R=0.7$ as used by \citet{SCH06a}.
To investigate the dependence of the control on $R$, we display the domains of control in the $(K,R)$-plane for
different values of the phase ($\varphi=0$, $\pi/8$, $\pi/4$, and $3\pi/8$) and fixed time delay $\tau=26$ in
Fig.~\ref{fig:lk_k_r_domain_phi}. This value of $\tau$ was chosen based on the results from the generic model considered
by \citet{HOE05,DAH07}, where it was shown that the optimum time delay is given by
$\tau=T_{0}/2=\pi/\mathrm{Im}(\Lambda_0)$, where $\Lambda_0$ denotes the eigenvalue of the uncontrolled system. In the
LK model, the imaginary part of the eigenvalues in the uncontrolled system is given by Eq.~(\ref{eqn:EIGENVALUES_LK})
with $K=0$, i.e., $\mathrm{Im}(\Lambda_0) =  \sqrt{4\Omega_{0}^{2}-\Gamma^{2}}$. This leads to an optimum time delay:
\begin{equation}
 \tau_{opt} = \frac{\pi}{\sqrt{4\Omega_{0}^{2}-\Gamma^{2}}} ,
 \label{eqn:tau_opt}
\end{equation}
which yields for our parameters $\tau_{opt} \approx 26$.

Now, in Fig.~\ref{fig:lk_k_r_domain_phi}, it can be seen that the domain of control in the $(K,R)$-plane has maximum
size for $\varphi=0$ for this choice of the time delay $\tau$. [See panel Fig.~\ref{fig:lk_k_r_domain_phi}(a).] For
increasing phase, the domain of control shrinks while moving to the upper right. Stability is then only achieved in a
small region
at large values of $K$ and $R$. [See panels Fig.~\ref{fig:lk_k_r_domain_phi}(b) to (d).]

\begin{figure}[ht]
	\begin{center}
		\includegraphics[width=0.975\linewidth]{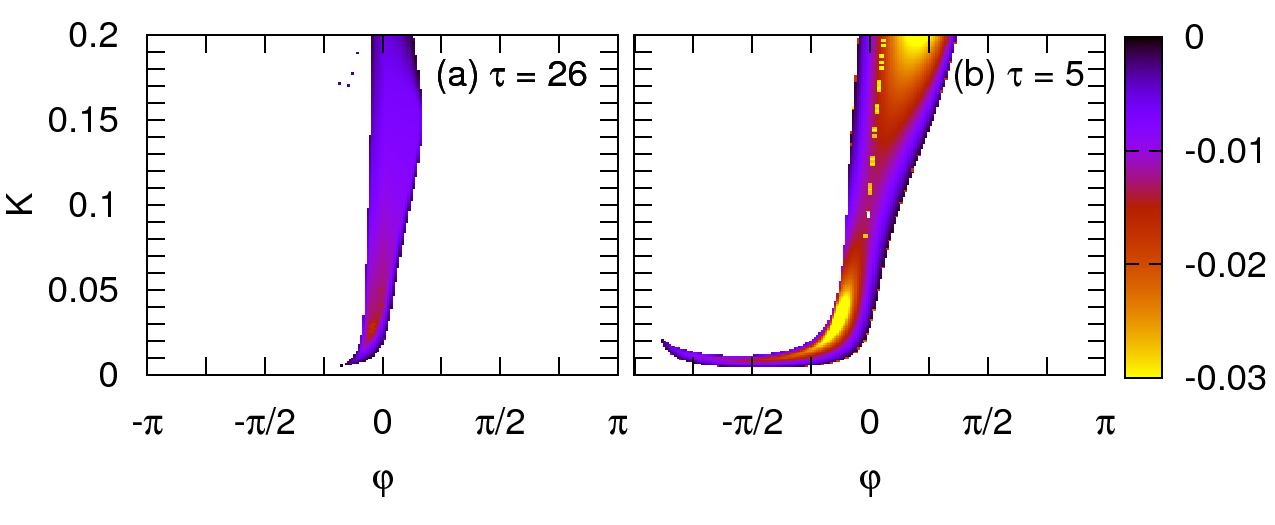}
	\end{center}
\caption{Domain of control in the $(K,\varphi)$-plane for different values of $\tau$ and fixed
$\delta=0$. The greyscale (color code) denotes the largest real part $\mathrm{Re}(\Lambda)$ of the eigenvalues
$\Lambda$; only negative values are plotted. Panels (a) and (b) correspond to $\tau=26$ and $5$, respectively. Other
parameters as in Fig.~\ref{fig:lk_k_tau_domain_phi}  \citep{DAH08b}.}
\label{fig:lk_k_phi_domain}
\end{figure}
To investigate the dependence of the domain of control on the choice of the phase $\varphi$ further, we consider another
two-dimensional projection of the five-dimensional control-parameter space parameterized by feedback gain $K$ and the
feedback phase $\varphi$. This section is depicted in  Fig.~\ref{fig:lk_k_phi_domain} for two different values of the
time
delay and fixed $\delta=0$. In panel (a), the time delay is chosen as $26$, which is the optimum $\tau$ according to
Eq.~(\ref{eqn:tau_opt}). Here, it can be seen that the optimum phase is located at slightly negative values for small
values of the feedback gain up to $K\approx 0.05$. Increasing $K$, the optimum phase changes its sign and is now located
at small positive values of $\varphi$. For the case of $\tau=5$, which is depicted in panel (b), stability is overall
enhanced drastically. The yellow areas, denoting regions of optimum stability, are located at negative
$\varphi$ for small $K$ up to $K \approx 0.1$. Control is possible even for a small value of $\varphi$ below
$-\pi/2$, if the feedback gain is tuned exactly to the small range of $K \approx 0.01$. For larger feedback gain with
$K>0.1$, the optimum value of $\varphi$ is located at positive values. The region of optimum stability is located at
large values of $K$ around $0.2$. The shape of the control domain in Fig.~\ref{fig:lk_k_phi_domain} is markedly
different from that in the generic normal form model (see Figs.~6(c) and 7(c) in Ref.~\citep{DAH07}), but appears to be
in line with full device simulations within a travelling wave model \citep{WUE07}.

To summarize, we have focused on simple rate equation models of neural
systems and lasers, which contain a time delay.  In these systems, analytical
results are important to complement computer simulations to determine whether
these systems fall into categories defined by the nature of their bifurcations.
This way, we have gained a more general understanding of the principles of
dynamical patterns and their control. Delay dynamics can, on one hand, add to
the dynamic complexity and generate bifurcations, and, on the other hand,
stabilize unstable states and suppress complex dynamics. Even in the most
generic neural systems described by the FitzHugh-Nagumo equation, when coupled
in a minimum network of two nodes with delayed coupling and delayed
self-feedback, we found complex scenarios of synchronized neuronal dynamics
with in-phase or antiphase oscillations, bursting patterns, and amplitude
death. In the case of the multisection semiconductor laser with an internal
passive dispersive reflector and external optical feedback described by the
modified Lang-Kobayashi model, we found that time-delayed self-feedback can
stabilize the laser. In the regime above a supercritical Hopf bifurcation,
where the fixed point in the uncontrolled system is unstable, unwanted
intensity pulsations can be suppressed by time-delayed feedback control.

\begin{ack}
This work was supaported by Deutsche Forschungsgemeinschaft in the framework of Sfb 555. 
\end{ack}


\end{document}